\documentclass[12pt,preprint]{aastex}
\usepackage{graphicx}
\usepackage{emulateapj5}           % emulates original format
\slugcomment{Accepted by the Astrophysical Journal}

\shorttitle{SMA sub-mm observations of IRAS\,18089-1732}
\shortauthors{Beuther et al.}

\begin{document}


\title{Testing the massive disk scenario for IRAS\,18089-1732}


\author{H.~Beuther, Q. Zhang, T.K. Sridharan, Y. Chen}

\affil{Harvard-Smithsonian Center for Astrophysics, 60 Garden Street, Cambridge, MA 02138, USA}

\email{hbeuther@cfa.harvard.edu}

%\and

%\author{R. J. Hanisch\altaffilmark{5}}
%\affil{Space Telescope Science Institute, Baltimore, MD 21218}


%\altaffiltext{1}{Visiting Astronomer, Cerro Tololo Inter-American Observatory.
%CTIO is operated by AURA, Inc.\ under contract to the National Science
%\altaffiltext{2}{Society of Fellows, Harvard University.}
%\altaffiltext{3}{present address: Center for Astrophysics,
%    60 Garden Street, Cambridge, MA 02138}
%\altaffiltext{4}{Visiting Programmer, Space Telescope Science Institute}
%\altaffiltext{5}{Patron, Alonso's Bar and Grill}


\begin{abstract} 

Investigating in more detail the previously suggested massive disk
scenario for the High-Mass Protostellar Object IRAS\,18089-1732, we
observed the source in the 860\,$\mu$m band with the Submillimeter
Array in various spectral lines and the submm continuum emission at
(sub-)arcsecond spatial resolution. Fifty spectral lines from eighteen
different species spanning upper level energy states between 17 and
747\,K were detected.  One of the assumed best tracers for massive
disks, CH$_3$CN, is optically thick and does not allow a further disk
investigation. However, the complex molecule HCOOCH$_3$ appears
optically thin and exhibits a velocity shift across the central
core perpendicular to the emanating outflow. This signature is
comparable to well-known low-mass disks and confirms the detection of
a massive rotating structure likely associated with the central
accretion disk. Assuming equilibrium between centrifugal and
gravitational force, the estimated mass for this rotating structure is
16/(sin$^2(i)$)\,M$_{\odot}$ (with $i$ the unknown inclination angle),
of the same order as the gas mass derived from the continuum
emission. Therefore, in contrast to low-mass disks, a considerable
amount of the central gas mass is associated with the rotation,
implying that massive disks may not be Keplerian. A temperature
estimate based on the CH$_3$CN(19--18) K-ladder results in $\sim
350$\,K, thus a hot core has already formed in this region.  An
analysis of the submm continuum emission reveals even at this high
spatial resolution only one isolated massive dust core without any
detectable companions down to mass limits between 0.2 and
3\,M$_{\odot}$ (depending on the assumed temperature). Potential
implications for the massive cluster formation are discussed. The
radial intensity distribution of the core is highly non-symmetric,
outlining the difficulties of density structure investigations based
on radial averaging.

\end{abstract}


%% subject header.  Objects should be in the appropriate "individual"
%% headers (e.g. quasars: individual, stars: individual, etc.) with the
%% additional provision that the total number of headers, including each
%% individual object, not exceed six.  The \objectname{} macro, and its
%% alias \object{}, is used to mark each object.  The macro takes the object
%% name as its primary argument.  This name will appear in the paper
%% and serve as the link's anchor in the electronic edition if the name
%% is recognized by the data centers.  The macro also takes an optional
%% argument in parentheses in cases where the data center identification
%% differs from what is to be printed in the paper.


\section{Introduction}
\label{intro}

The phenomenon of massive protostellar disks is one of the central
questions in high-mass star formation. Indirect indicators for the
existence of massive disks has been accumulated over recent years but
unambiguous evidence for massive disks is still
missing. High-spatial-resolution observations of massive molecular
outflows revealed that they could be as collimated as their low-mass
counterparts (e.g.,
\citealt{beuther2002d,gibb2003,garay2004,davis2004}) indicating some
underlying protostar-disk interaction as the driving
mechanism. Furthermore, there have been various approaches to study
massive disks in a more direct way. Infrared and mm continuum
observations find interesting disk-candidates (e.g.,
\citealt{shepherd2000,chini2004}), but they lack the necessary
velocity information to provide kinematic proof of potential accretion
disks. Contrary to this, Class II CH$_3$OH and H$_2$O maser emission
found disk-like kinematic signatures in a few cases (e.g.,
\citealt{torrelles1996,pestalozzi2004}), but since maser emission is
very selective and requires special conditions such studies do not
allow a more statistical investigation of massive disk properties. The
most promising approach to find and investigate disks in high-mass
star formation appears to be the observation of thermal line emission
which should be sensitive to the gas properties and kinematics of the
disks. Various investigations have been directed this way: one of the
first massive disk studies were the CH$_3$CN observations toward
IRAS\,20126+4104 which found a velocity gradient perpendicular to the
outflow \citep{cesaroni1997,cesaroni1999}. More recently,
\citet{beltran2004} also used CH$_3$CN observations to analyze
potential rotating structures toward two even more massive high-mass
star-forming regions, and again they found velocity gradients
perpendicular to observed molecular outflows. A different approach was
followed by \citet{zhang1998a,zhang2002} who investigated the disks of
IRAS\,20126+4104 and AFGL5142 by the means of NH$_3$
observations. While these approaches are promising, there is still no
unambiguous evidence for disks in high-mass star formation. Even more
troubling is the fact that we still do not know which is the best
molecular transition to study massive disks, CH$_3$CN seems to be a
promising candidate and NH$_3$ works in some cases as well, but we
also know a number of sources where the NH$_3$ observations apparently
do not trace the disks (Zhang, priv.~comm.). One of the challenges of
the coming years is to establish a reliable procedure to find massive
disks and study their properties, which is important to constrain any
theory of high-mass star formation.

With this problem in mind, we began high-spatial-resolution
observations of the HMPO IRAS\,18089-1732 in various molecular lines
with the Submillimeter Array (SMA\footnote{The Submillimeter Array is
a joint project between the Smithsonian Astrophysical Observatory and
the Academia Sinica Institute of Astronomy and Astrophysics, and is
funded by the Smithsonian Institution and the Academia Sinica.}) last
year \citep{beuther2004a,beuther2004b}. This source at a distance of
$\sim 3.6$\,kpc with a luminosity of $\sim 10^{4.5}$\,L$_{\odot}$ is
part of a sample of 69 HMPOs intensively studied over recent years by
\citet{sridha,beuther2002a,beuther2002b,beuther2002c}. The single-dish 
millimeter continuum observations reveal a gas mass of $\sim
2000$\,M$_{\odot}$, and H$_2$O and Class II CH$_3$OH maser emission as
well as a weak cm continuum source (0.9\,mJy) has been detected at the
core center. Previous SMA observations found an outflow in north-south
direction and HCOOCH$_3$ emission centered at the mm continuum peak
exhibiting a velocity gradient approximately perpendicular to the
outflow \citep{beuther2004b}. The latter observations were interpreted
as indicative of a possible rotating disk-like structure around the
central protostar which could be influenced by the outflow as well as
infall. The SiO outflow emission is slightly offset from the mm
continuum peak ($\sim 1''$), allowing to speculate about a potential
secondary source undetected so far. Observations of a series of other
molecular transitions revealed that many of them trace to some degree
the outflow as well, and only a few remain confined to the core
center, CH$_3$OH being one of the latter species
\citep{beuther2004a}. However, some of the typical disk-tracers like
CH$_3$CN were not covered by the spectral setup. Furthermore, the
spatial resolution was marginal for our purposes ($2.7''\times 1.7''$
corresponding to approximately 7000\,AU at the given
distance). Therefore, we followed up on the previous observations via
higher spatial resolution ($\leq 1''$) observation covering a CH$_3$CN
series as well. These observations are intended to get a better
understanding of the potential massive accretion disk as well as to
investigate which molecular lines are best suited for future massive
disk studies.

\section{Observations}
\label{obs}

We observed the HMPO IRAS\,18089-1732 in 3 nights during the summer of
2004 with the SMA in two configurations covering unprojected baseline
lengths between 20 and 340\,m. The phase center was
R.A.~18:11:51.4/Dec.~-17:31:28.5 (J2000) and the tuning frequency
349.45\,GHz in the upper sideband ($v_{\rm{lsr}}=-33.8$). Two out of
the three nights the weather was good with a zenith opacity of 0.18 at
348\,GHz~-- scaled from the 225\,GHz opacity measurement of the NRAO
tipping radiometer operated by the CSO~. In the third night, the
weather was worse with $\tau(348\,\rm{GHz})\sim 0.38$, the latter data
were weighted considerably lower in the data reduction. The spectral
resolution over the whole bandpass of 4\,GHz (upper + lower sideband)
was 0.8125\,kHz corresponding to a velocity resolution of $\sim
0.7$\,km/s. The initial flagging and calibration was done in the IDL
superset MIR originally developed for the Owens Valley Radio
Observatory \citep{scoville1993} and adapted for the SMA\footnote{The
MIR cookbook by Chunhua Qi can be found at
http://cfa-www.harvard.edu/$\sim$cqi/mircook.html.}. The imaging and data
analysis was conducted in MIRIAD. The passband calibration was derived
from a variety of different sources for each track including Jupiter,
Neptune, Uranus, Callisto, Mars and 3c279 (for the largest
baselines). Flux calibration was conducted with observations of
Uranus, Callisto and 3c279 (again for the largest baselines). The flux
density scale is estimated to be accurate within 15\%. For the phase
and amplitude calibration we used regularly interleaved observations
of the quasar NRAO530 (about $10^{\circ}$ from the target source) with
an approximate flux of $1.5$\,Jy at the given frequency.  We applied
different weightings for the continuum and line data resulting in
synthesized beams for the continuum of $1.1''\times
0.5''\,(\rm{P.A.}\,1.3^{\circ})$ and for the lines of $1.3''\times
0.9''\,(\rm{P.A.}\,2.2^{\circ})$. The 1$\sigma$ rms of the continuum
image is 35\,mJy.

\section{Results}

\subsection{860\,$\mu$m continuum emission}
\label{continuumsection}

\begin{figure}
\includegraphics[angle=-90,width=9cm]{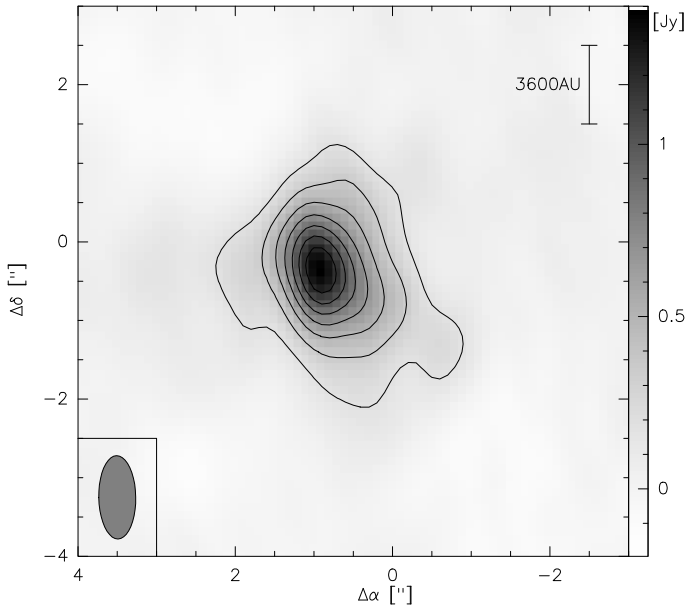}
\caption{860\,$\mu$m continuum image of IRAS\,18089-1732. The contours
range from 175 to 1400\,mJy in 175\,mJy steps. The beam ($1.1''\times
0.5''$\,P.A. 1$^{\circ}$) is shown at the bottom-left. The deconvolved
source size is $\sim 3000$\,AU (using the 50\% contour as the observed
image size). The axis offsets are in arcseconds from the phase center
(\S \ref{obs}).}
\label{continuum}
\end{figure}

\begin{figure}
\includegraphics[angle=-90,width=8cm]{f2a.ps}
\includegraphics[angle=-90,width=8cm]{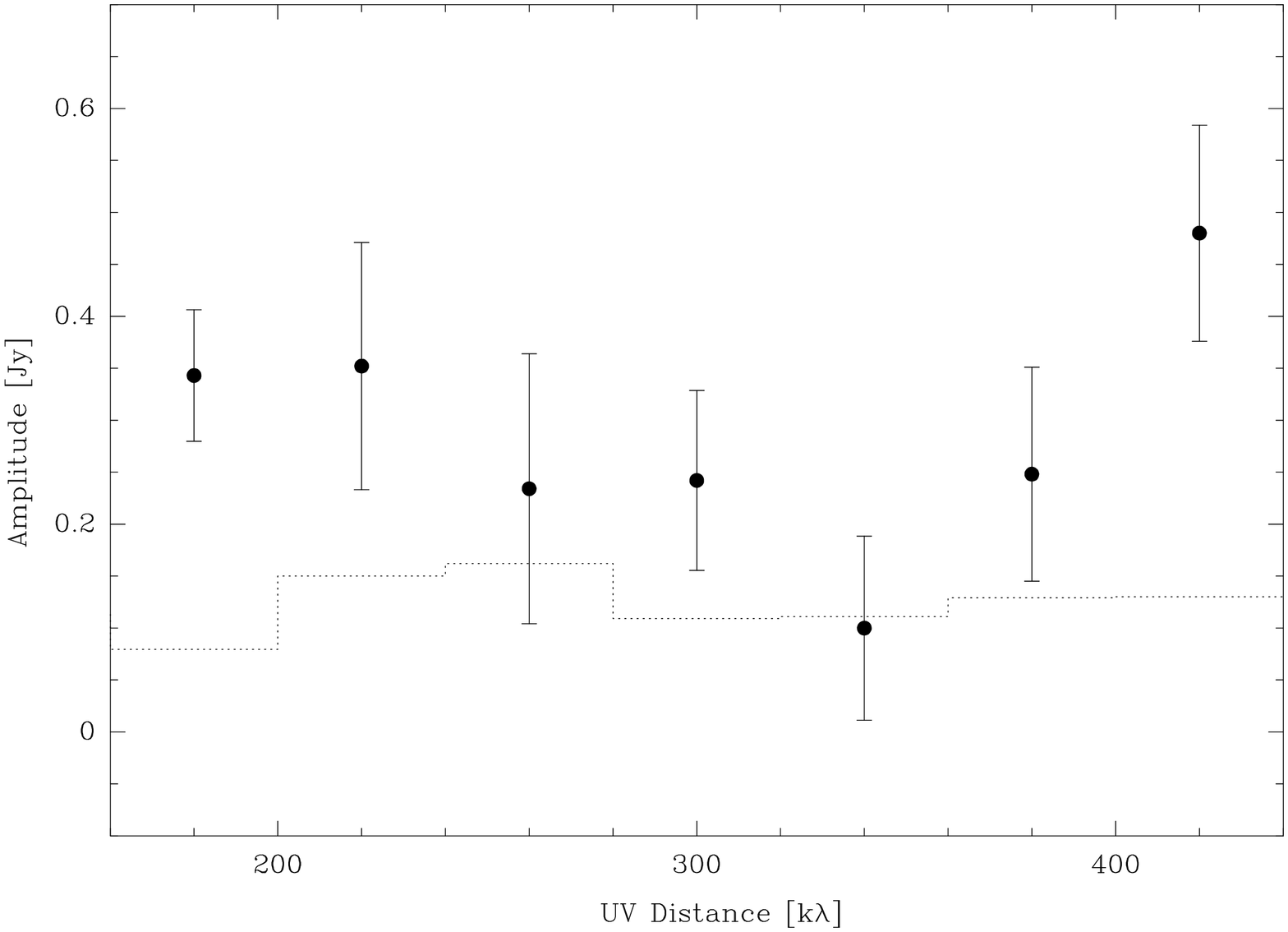}
\caption{Amplitude versus uv-distance plots of the continuum data
presented in Figure \ref{continuum}. The left panel shows the whole
dataset in 5\,k$\lambda$ bins, the right panel presents only the long
baseline data in larger 40\,k$\lambda$ bins. The dotted lines show
the expected amplitude for zero signal with the given statistical
errors.}
\label{amp_uvdistance}
\end{figure}

Figure \ref{continuum} shows the continuum image with $1.1''\times
0.5''$. In spite of some structure, it remains a single source and
does not split up into multiples. Two corresponding amplitude versus
uv-distance plots are shown in Figure \ref{amp_uvdistance}.  While the
shortest baselines show a halo component, the range between 50 and
90\,k$\lambda$ appears rather flat consistent with the previously
modeled point source \citep{su2004b}. However, from 90\,k$\lambda$
onward the flux densities decrease further, indicating additional
sub-structure on the corresponding spatial scales. Beyond
200\,k$\lambda$, the S/N ratio is low, and in the 5\,k$\lambda$ bin
plot (Fig.~\ref{amp_uvdistance}\,left) one cannot differentiate the
emission from statistical noise. Averaging the data into larger
40\,k$\lambda$ bins to increase the S/N ratio
(Fig.~\ref{amp_uvdistance}\,right), we can discern emission above the
noise. The large baseline data are consistent with a point source of
approximate flux density $\sim 0.3$\,Jy. The integrated flux density
of the shorter baseline halo component can be fitted with a Gaussian
to $\sim 6.0$\,Jy.

\begin{figure}
\includegraphics[angle=-90,width=12cm]{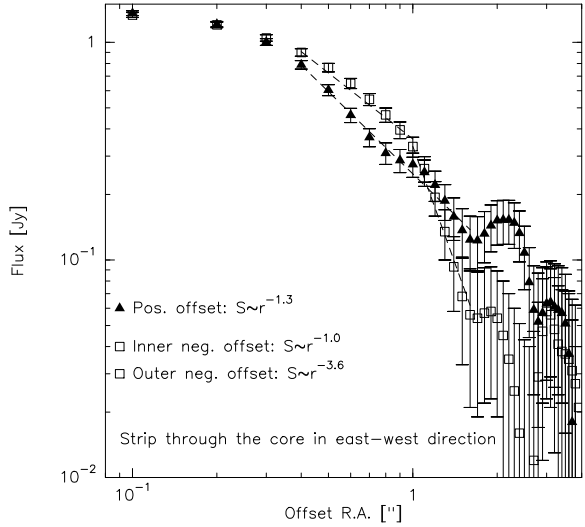}
\caption{Continuum flux versus peak-offset plot. The data are taken
along a strip through the continuum peak in east-west direction~-- the
direction with the smallest beam-size of $0.5''$. The filled triangles
mark positive offsets and the open squares negative offsets. The
dashed lines show power-law fits to subsets of the data, outlining
that there is significant variation even along this singular strip.}
\label{strip}
\end{figure}

As a continuation of our radial profile study toward this source with
single-dish 1.2\,mm continuum observations (with the MAMBO array at a
spatial resolution of $\sim 11''$, \citealt{beuther2002a}), we now
attempted to derive radial intensity profiles from these higher
spatial resolution data. Since the beam is elliptical, we extracted a
strip in the east-west direction (smallest beam size of $0.5''$) and
derived the profiles in the directions of positive and negative
R.A. offsets, respectively. The results and some fits to the data are
shown in Figure \ref{strip}. Obviously, the intensity profiles do vary
significantly in the two directions with power-law indices between
$-1.0$ and $-3.6$. Thus, the typical approach of circular averaging is
not useful here. Nevertheless, to better quantify the errors often
introduced by circular averaging, we restored the image with a
circular beam of $\sim 0.75''$ and derived the intensity profile using
circular averaging. Not surprisingly, the derived power law index (not
shown) is between the ones shown in Figure \ref{strip} with an index
of $-2.1$. One result to be kept in mind is that circular averaging has to
be done cautiously, at least at this spatial resolution.

Table \ref{submm} shows the submm continuum peak $S_{\rm{peak}}$ and
integrated intensity $S_{\rm{int}}$. Comparing these values with the
previously derived continuum flux densities \citep{beuther2004b}, we
find nearly the same peak intensity but about three times the
integrated flux densities. The synthesized beam of the new dataset is
a bit smaller than that of last years observations ($1.1''\times
0.5''$ vs. $1.4''\times 0.9''$). However, as shown in
Fig.~\ref{amp_uvdistance}, data $>150$\,k$\lambda$ do not contribute
strongly to the measurable flux densities, and thus finding similar
peak fluxes densities in both datasets is reassuring that the
calibration of the SMA is reliable. The fact that we observe
significantly higher integrated flux densities in the new dataset is
due to the better uv-coverage we have this time. Due to this better
uv-coverage we suffer less from missing short spacings and recover
more of the extended halo emission. Nevertheless, measuring the
integrated flux densities of the 850\,$\mu$m single-dish SCUBA
observations \citep{williams2004} in an aperture with a radius of
$\sim 30''$, we get an integrated flux density of $\sim 30$\,Jy. Thus,
we only recover about 20\% of the single-dish flux densities with the
SMA, the remaining 80\% mostly reside in a larger-scale halo filtered
out by the interferometer, and a smaller fraction of the integrated
flux is lost because of the negative side-lobe emission.

Following \citet{hildebrand1983}, assuming the submm continuum
emission to be due to optically thin dust emission, we calculate
the gas mass and column density of the central core. The procedure
and equations we use are outlined in \citet{beuther2002a}. Based on
the spectral index analysis from \citet{beuther2004b} we assume a dust
opacity index $\beta =1$. Furthermore, we calculate the gas mass $M$
on the one hand using a typical hot core temperature of 100\,K to
better compare it with the results by \citet{beuther2004b}, and on the
other hand using the temperature of 350\,K derived from the CH$_3$CN
observations (see \S\ref{ch3cn}). As discussed in
\citep{beuther2002a}, the errors are dominated by systematics, e.g.,
the exact knowledge of the dust opacity index $\beta$ or the
temperature. While we take different temperatures into account the
other parameters are more difficult to acknowledge and we estimate the
masses and column densities to be correct within a factor 5.

Assuming 100\,K we get a gas mass of 45\,M$_{\odot}$ close to the
38\,M$_{\odot}$ derived last year from the 1.3\,mm continuum
observations \citep{beuther2004b}, whereas using the 350\,K results in
a lower gas mas of 12\,M$_{\odot}$. The values will be compared below
with the masses derived from the kinematic HCOOCH$_3$ observations
(see \S\ref{hcooch3}). Regarding the gas column density $N$, both
temperatures result in values of the order $10^{24}$\,cm$^{-2}$, close
to the column densities previously derived from the lower spatial
resolution 1.3\,mm data \citep{beuther2004b}. Assuming a relation
between column density and extinction of
$A_v(\rm{mag})=N/(10^{21}\,cm^{-2})$ we get a visual extinction of the
order 1000.

\subsection{Spectral line data}

\begin{figure}
\begin{center}
\includegraphics[angle=-90,width=10cm]{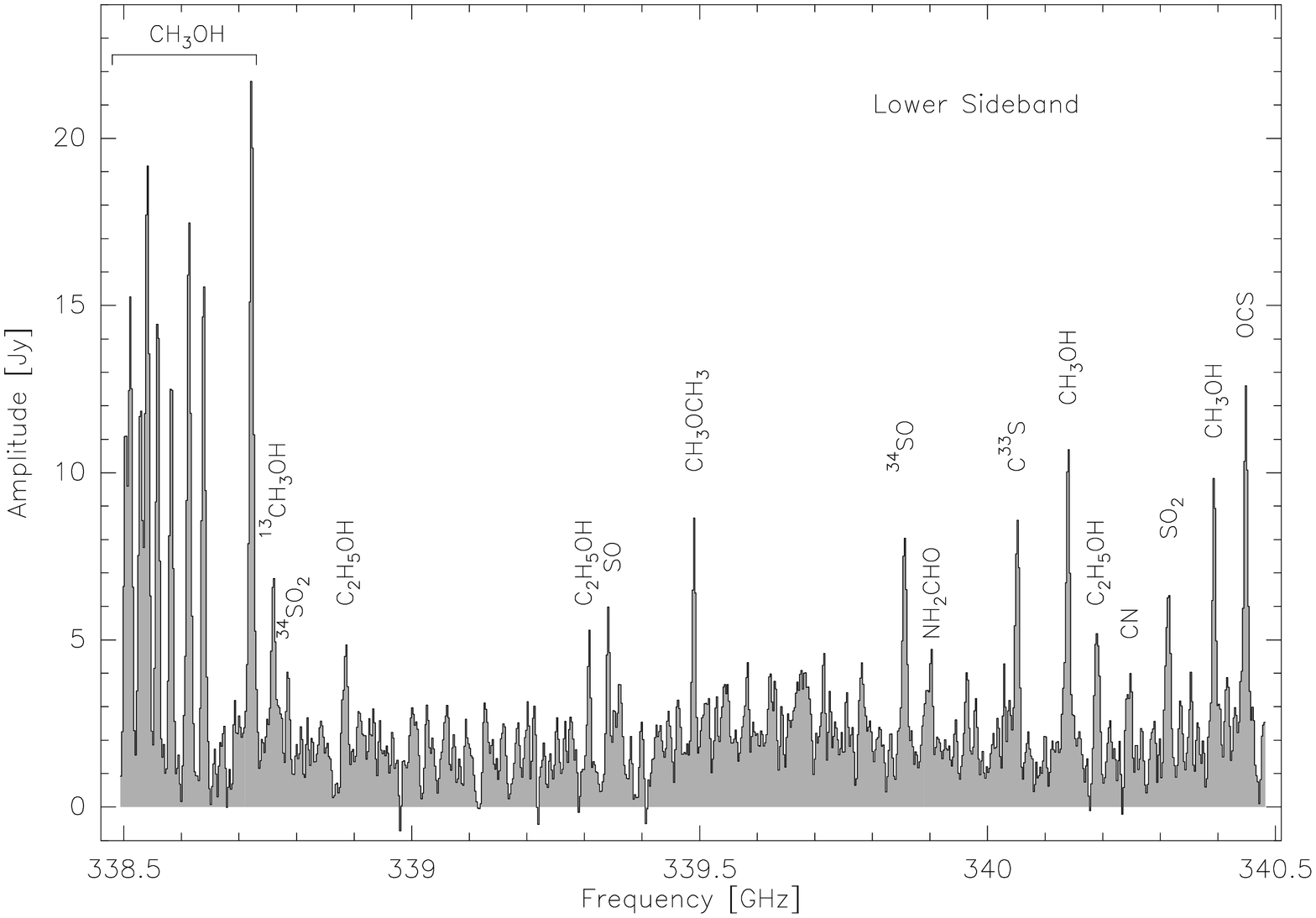}\\
\includegraphics[angle=-90,width=10cm]{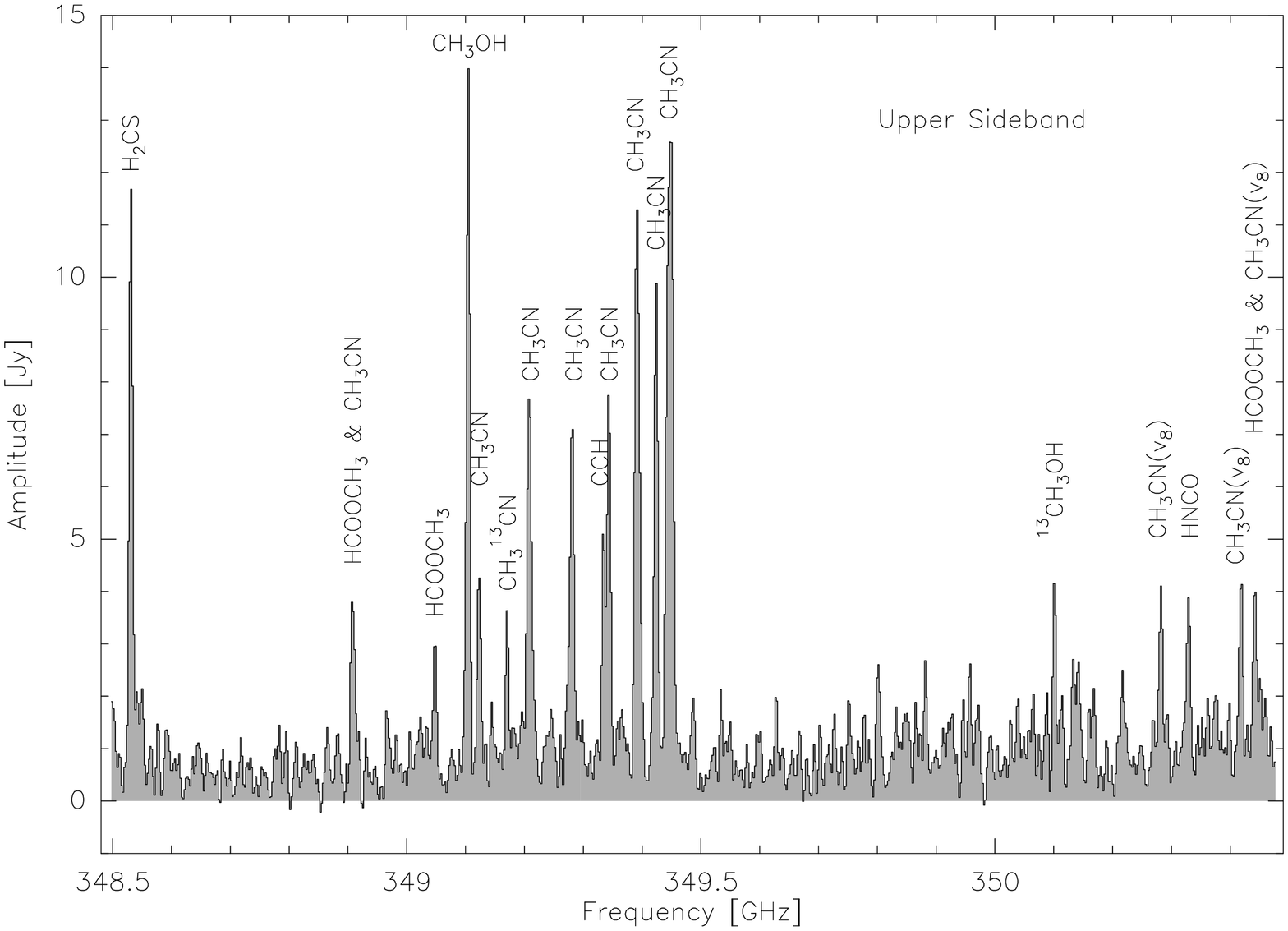}
\end{center}
\caption{Vector-averaged spectra on a short baseline of 26\,m. Most 
molecular lines from Table \ref{lines} are marked.}
\label{spectra_uv}
\end{figure}

\begin{figure}
\includegraphics[angle=-90,width=16.5cm]{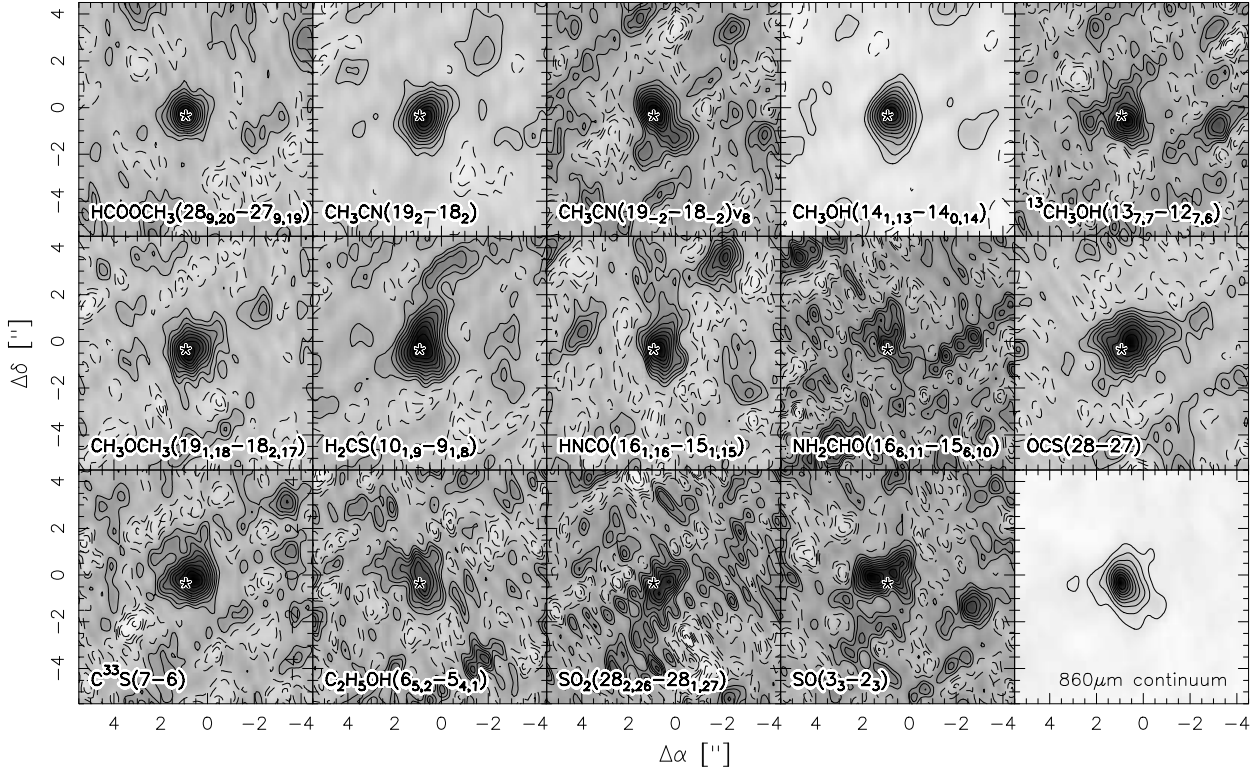}
\caption{Integrated images of all observed molecular species. The
contour levels are chosen for each image from $\pm 10$ to $\pm 90$\%
of its peak emission. The star marks the submm continuum peak position
(Fig.~\ref{continuum}).}
\label{images}
\end{figure}

Figure \ref{spectra_uv} shows vector-averaged spectra of the complete
spectral setup taken on a short baseline of 26\,m, and Table
\ref{lines} lists the detected lines. Line identification were
conducted with the 350\,GHz spectral line survey by
\citet{schilke1997b} and the molecular line databases JPL
\citep{poynter1985} and CDMS \citep{mueller2001}. Altogether 50
spectral lines from 18 species are observed (we consider different
isotopomers as different species: CH$_3$OH, $^{13}$CH$_3$OH, SO$_2$,
$^{34}$SO$_2$, SO, $^{34}$SO, C$_2$H$_5$OH, CH$_3$OCH$_3$, NH$_2$CHO,
C$^{33}$S, CN, OCS, H$_2$CS, HCOOCH$_3$, CH$_3$CN, CH$_3^{13}$CN, CCH,
HNCO). Interestingly, within this single spectral setup we detect
molecular lines spanning upper state energy levels $E_{\rm{upper}}$
from 17 to 747\,K (Table \ref{lines}). Fourteen species are
sufficiently strong to be imaged.  Figure \ref{images} shows the
integrated images of these species, and only HCOOCH$_3$ does peak
exactly on the continuum image. A more detailed analysis of the
HCOOCH$_3$ data is given in \S\ref{hcooch3}. The lines we were
originally mainly interested in, the CH$_3$CN$(19_k-18_k)$ K-ladder, appear
to be optically thick, and do not properly trace the most central core
(see \S\ref{ch3cn}). 

Comparing the other molecular lines with the previously observed SiO
outflow emission leaving the core in the northern direction
\citep{beuther2004b}, some lines show contributions from the outflow
as well, most strongly H$_2$CS. Most of the other species appear to
peak close to the continuum emission~-- usually within $\leq 1''$
corresponding to $\leq 3600$\,AU~-- but none directly toward
it. Interestingly, for most species this offset is toward the west and
north-west, similar to the peculiar offset of the SiO outflow axis
\citep{beuther2004b}. Since we do not detect any companion source in the 
submm continuum data (see also \S \ref{discussion_cont}), it is
unlikely that a secondary source is causing this offset. It appears
more plausible that shock interactions of the outflow with the ambient
molecular core enhance the molecular abundances and thus at least
partly cause the observed offsets. It is difficult to quantify the
relative contributions from excitation, chemistry, opacity, outflow
and other kinematic effects to the morphological differences between
the various species, and likely an interplay of these effects has to
be taken into account to explain all observational features.

Comparing the spatial extent of the 50\% contour levels from some of
the molecules in Table \ref{area} (we selected just the strong ones
with reasonable S/N ratio), we find that HCOOCH$_3$ is confined most
strongly, approximately within the same region as the 860\,$\mu$m
continuum emission, whereas all other molecules, even density tracers
like CH$_3$CN/CH$_3$OH/C$^{33}$S, are more extended. With HCOOCH$_3$ peaking
exactly toward the submm continuum peak and showing the smallest
spatial extend, it appears to be one of the best dense gas tracers
within this mini-line-survey.

\subsubsection{CH$_3$CN$(19_k-18_k)$}
\label{ch3cn}

\begin{figure}
\includegraphics[angle=-90,width=10cm]{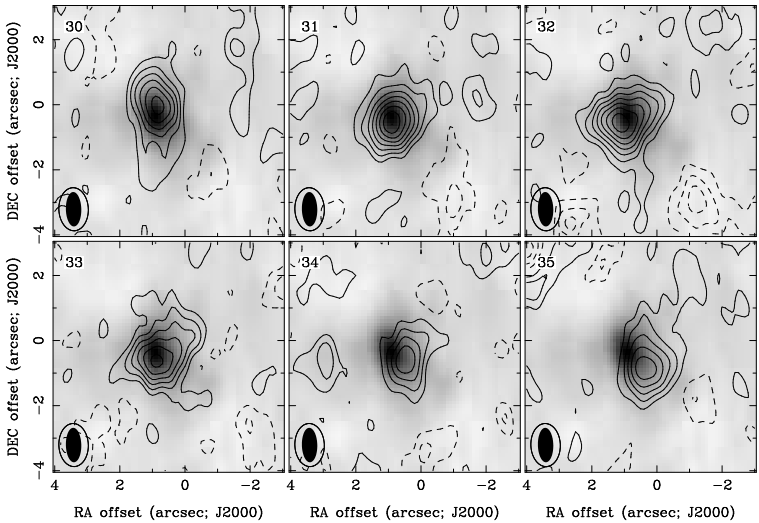}
\caption{Channel map of CH$_3$CN($19_2-18_2$). The grey-scale shows
the 860\,$\mu$m continuum image and the contours present the CH$_3$CN
data. Positive contour levels (full lines) are 0.67 to 4.69\,Jy
(0.67\,Jy steps), and negative contours range from $-0.67$ to
$-2.01$\,Jy also in $-0.67$\,Jy steps.}
\label{channel_ch3cn}
\end{figure}

All CH$_3$CN$(19_k-18_k)$ velocity channels peak offset from the mm
continuum peak, and we cannot distinguish a clear velocity structure
(Fig.~\ref{channel_ch3cn}). The picture is similar for all 8 detected
CH$_3$CN$(19_k-18_k)$ K-lines (K=0,...,7). While physical reasons like
the above mentioned outflow contributions can account for some of the
offsets, the rather random-like distribution of offsets in this
channel map is indicative of high optical depths throughout the whole
K-ladder (see also the discussion about the spectral fits below).
Thus, the CH$_3$CN$(19_k-18_k)$ lines appear not to be suited well as
a disk tracer for this source.

\begin{figure}
\includegraphics[angle=-90,width=12cm]{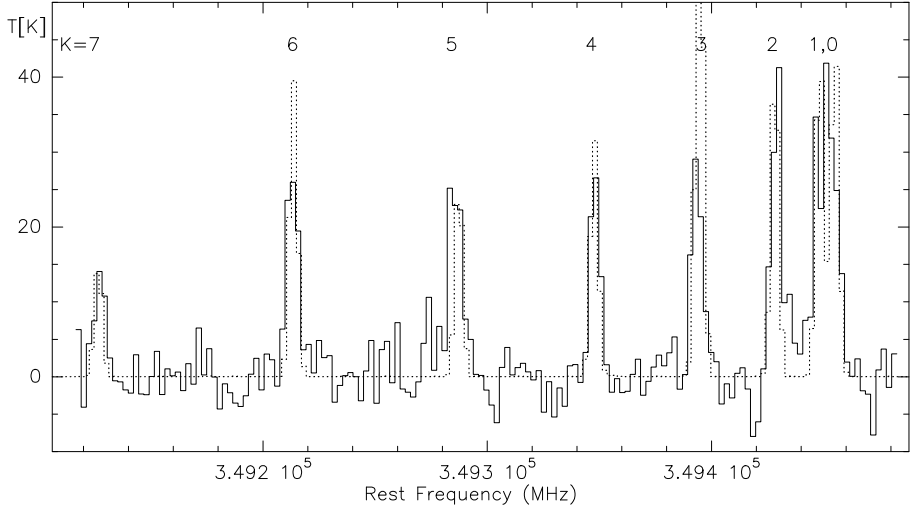}
\caption{The full line shows the CH$_3$CN$(19_k-18_k)$ spectrum taken 
toward the submm continuum peak after imaging the lines. The dotted
line presents the model spectrum at 350\,K with a column density of
$1.4\times 10^{15}$\,cm$^{-2}$. The K-levels are indicated above each
line.}
\label{ch3cn_fit}
\end{figure}

Nevertheless, we can use the CH$_3$CN K-ladder to get a
temperature estimate of the core. For this purpose, we used the XCLASS
superset of the CLASS program (Schilke, priv.~comm.). This
software-package uses the line catalogs from JPL and CDMS,
\citealt{poynter1985,mueller2001}). We tried to fit the whole CH$_3$CN
spectrum simultaneously assuming optically thin emission in local
thermodynamic equilibrium. Figure \ref{ch3cn_fit} presents the whole
CH$_3$CN$(19_k-18_k)$ spectrum toward the submm peak position and a
model spectrum using a temperature of 350\,K and a CH$_3$CN column
density of $1.4\times 10^{15}$\,cm$^{-2}$. This model fits the data
good for the K=0,1,2,4,5,7 lines but bad for K=3,6. The ortho-CH$_3$CN
(A-species) lines K=3,6 have double the statistical weight than the
other observed lines, whereas the observed line strengths of the
K=3,4,5,6 lines (including para and ortho species) are similar. The
observed line ratio close to unity between ortho- and para-CH$_3$CN
lines is additional evidence for the large optical depths of the
lines. In spite of these difficulties, the temperature of 350\,K is
consistent with our previous temperature estimate based on HCOOCH$_3$
observations \citep{beuther2004a} and thus gives a reasonable
temperature for the central forming hot molecular core.

\subsubsection{Disk signature in HCOOCH$_3$}
\label{hcooch3}

In contrast to the CH$_3$CN emission, HCOOCH$_3$ traces the central
dust emission well and thus appears to be likely optically thin.
Unfortunately, the spectral line at $\sim 348.91$\,GHz is weaker than
most of the CH$_3$CN lines, and it is a blend of two HCOOCH$_3$ lines
and the CH$_3$CN($19_9-18_9$) line (see Table \ref{lines}). However,
the energy levels of the HCOOCH$_3$ lines are at $\sim 294$\,K whereas
for CH$_3$CN($19_9-18_9$) it is at $\sim 745$\,K (Table
\ref{lines}). Therefore, the spectrum is dominated by the
HCOOCH$_3$ lines. Furthermore, using the rest-frequency of the
HCOOCH$_3$($28_{9,20}-27_{9,19}$)-E line at 348.9095\,GHz, the line
center is at 32\,km/s consistent with the line centers of other dense
gas tracers, e.g., CH$_3$CN. The frequency of the corresponding
HCOOCH$_3$($28_{9,20}-27_{9,19}$)-A line should be blue-shifted by
$\sim 4.7$\,km/s but we do not find a significant secondary peak
toward this velocity. Likely, there is underlying emission from this
A-type line, although at weaker intensities, but the spectrum is
dominated by the HCOOCH$_3$ line of the corresponding E-species. Thus,
it appears feasible to use this HCOOCH$_3$ line for a closer analysis
of the kinematics of the central core.

\begin{figure}
\includegraphics[angle=-90,width=16.5cm]{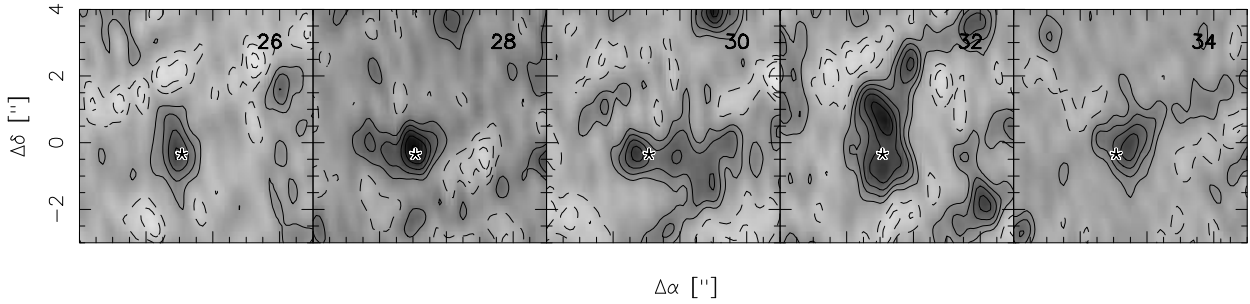}
\caption{Channel map of HCOOCH$_3$. The positive contour levels are in
solid lines from 0.35 to 1.75\,Jy (0.35\,Jy steps), the negative
contours are in dashed lines from -0.35 to -1.05\,Jy (0.35\,Jy steps)
The central velocities are marked at the top-right of each panel, and
the star pinpoints the submm continuum position
(Fig.~\ref{continuum}).}
\label{channel_hcooch3}
\end{figure}

\begin{figure}
\begin{center}
\includegraphics[angle=-90,width=10cm]{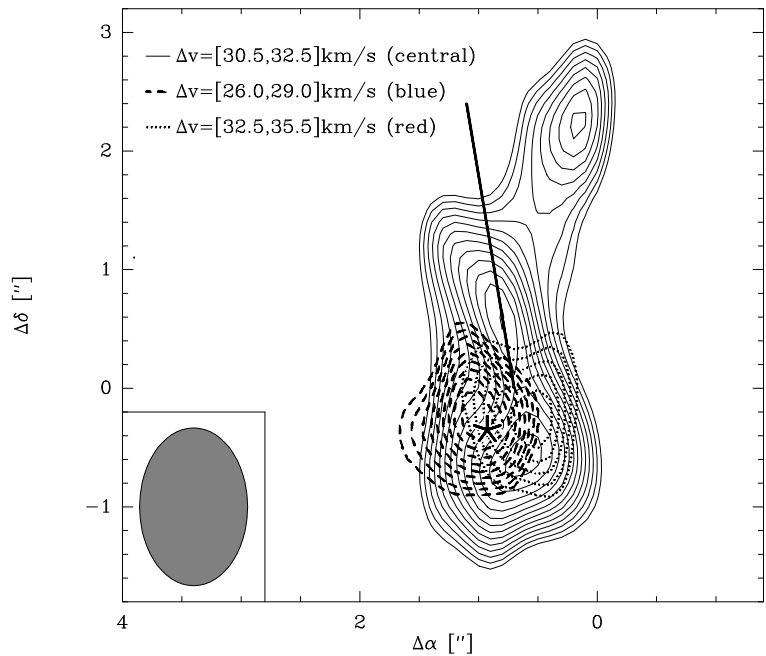}
\end{center}
\caption{Disk signature in HCOOCH$_3$. Dashed and dotted lines show
the blue and red disk emission, respectively. The full lines present
the emission at intermediate velocities. Again, the contour levels are
chosen from 10 to 90\% of the peak emission of each velocity bin
separately. The bold lines marks the outflow direction observed in
SiO(5--4) \citep{beuther2004b}.}
\label{disk}
\end{figure}

Figure \ref{channel_hcooch3} shows a channel map of this
HCOOCH$_3$($28_{9,20}-27_{9,19}$)-E line. Consistent with our previous
investigation \citep{beuther2004b}, we find a velocity shift in the
east-west direction over the core\footnote{We note that one cannot
directly compare the velocity channels of the previous HCOOCH$_3$
observation near 216.966\,GHz with our new data, because in last
year's analysis we set the rest-frequency to a central component of
the line blend whereas for the new data it is set to the strongest
component HCOOCH$_3$($28_{9,20}-27_{9,19}$)-E. This introduces
slightly different velocity ranges for both observations, but it does
not change the relative velocity shift over the source.} . This can be
depicted in the blue channels at 26 and 28\,km/s, which show compact
emission close to and a bit east of the mm continuum peak, and the red
channel at 34\,km/s, which shows compact emission west of the submm
continuum peak. The intermediate channels at 30 and 32\,km/s show
slightly more extended emission features. The fact that we see blue
emission over a little bit wider velocity range than red emission may
be due to the weaker HCOOCH$_3$($28_{9,20}-27_{9,19}$)-A line, which
is blue-shifted with regard to the main
HCOOCH$_3$($28_{9,20}-27_{9,19}$)-E line. The red emission of the
HCOOCH$_3$($28_{9,20}-27_{9,19}$)-A line would then be undetectable
because of the stronger ambient gas emission at 30 and 32\,km/s from
the HCOOCH$_3$($28_{9,20}-27_{9,19}$)-E line.  Figure \ref{disk}
presents an overlay of the HCOOCH$_3$ data binned into
blue/central/red components. Additionally, the outflow axis is
sketched as traced by the SiO(5--4) observations
\citep{beuther2004b}. With the rotation axis, defined by the
high-density tracer HCOOCH$_3$, being approximately perpendicular to
the outflow axis, these observations in IRAS\,18089-1732 show a disk
signature similar to those known from many low-mass sources (e.g., D
Tauri, \citealt{guilloteau1998}). This confirms at higher spatial
resolution our previously proposed disk scenario for this High-Mass
Protostellar Object. While the HCOOCH$_3$ observations at 216.966\,GHz
showed a P.A. of $\sim 55^{\circ}$ (from north), the axis of the new
data is closer to $90^{\circ}$. Considering the different energy
levels of the various HCOOOCH$_3$ transitions
($E_{\rm{upper}}(\rm{HCOOCH_3}(20_{2,19}-19_{0,19})) \sim 111$\,K \&
$E_{\rm{upper}}(\rm{HCOOCH_3}(28_{9,20}-27_{9,19})) \sim 294$\,K),
this implies that the lower-energy-level transitions are more strongly
affected by the infalling surrounding envelope. In contrast, the
higher-energy-lines appear to be better probes of the actual
underlying disk properties.

The deconvolved source size of the integrated HCOOCH$_3$ map in
Fig.~\ref{images} is $\sim 2050$\,AU (using the 50\% contour as the
observed image size), similar to the separation of the blue- and
red-shifted HCOOCH$_3$ peaks in Fig.~\ref{disk} of $\sim
1800$\,AU. With this source size, we cannot clearly distinguish
whether we observe the outer regions of a circumstellar disk or rather
a larger-scale rotating molecular envelope/torus which could feed the
inner accretion disk. Assuming equilibrium between the centrifugal and
gravitational forces at the outer radius of the disk the enclosed
dynamical mass can be estimated via:

\begin{eqnarray}
M_{\rm{disk}} & = & \frac{v^2r}{G} \label{eq1} \\
\Rightarrow M_{\rm{disk}}[\rm{M_{\odot}}] & = &  1.13\,10^{-3} \times v^2[\rm{km/s}]\times r[\rm{AU}] \label{eq2}
\end{eqnarray}

\noindent where $M_{\rm{disk}}$ is the disk mass, $r$ the disk radius,
in our case half the separation of the blue and red HCOOCH$_3$
emission peaks at most extreme blue and red velocities, and $v$ the
Half Width Zero Intensity (HWZI) of the spectral line. Equations
\ref{eq1} \&
\ref{eq2} have to be divided by sin$^2(i)$ where $i$ is the unknown
inclination angle between the disk plane and the line of
sight. Adopting the values $v\sim 4$\,km/s and $r\sim 900$\,AU, the
resulting disk mass $M_{\rm{disk}}$ is $\sim
16$/(sin$^2(i)$)\,M$_{\odot}$\footnote{The HWZI of the whole observed
line is $\sim 5$\,km/s. Since there is some blending between the
HCOOCH$_3$($28_{9,20}-27_{9,19}$) main E-type and weaker A-type line,
we adopt a slightly smaller value of HWZI=$\sim 4$\,km/s for the
HCOOCH$_3$($28_{9,20}-27_{9,19}$)-E transition.}. This value is close
to the 22/(sin$^2(i)$)\,M$_{\odot}$ value derived from our previous
lower resolution 1.3\,mm HCOOCH$_3$ observations
\citep{beuther2002b}. It also compares well to the 12\,M$_{\odot}$
core mass derived assuming 350\,K (in contrast to the estimated
45\,M$_{\odot}$ assuming 100\,K, see \S\ref{continuumsection} and
Table \ref{submm}). As already noticed by several studies (e.g.,
\citealt{cesaroni1997,zhang1998a,sandell2003,beltran2004,beuther2004b}),
in contrast to low-mass disks where the disk mass is usually
negligible compared with the core mass, a considerable amount of the
gas mass appears to be associated with large accretion disks and/or
rotating envelopes in the central regions of high-mass star formation.

\subsection{Varying spatial distributions of CH$_3$OH}
\label{ch3oh}

\begin{figure}
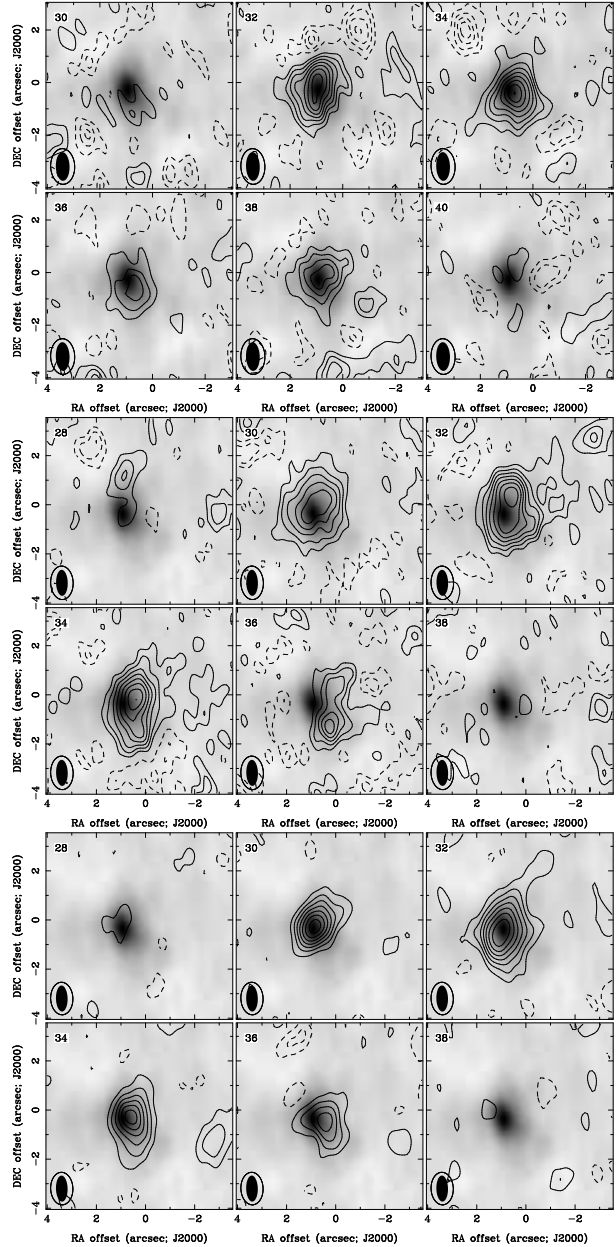

\begin{center}
\includegraphics[angle=-90,width=8cm]{f10a.ps}\\
\includegraphics[angle=-90,width=8cm]{f10b.ps}\\
\includegraphics[angle=-90,width=8cm]{f10c.ps}
\end{center}
\caption{Channel maps of the CH$_3$OH($2_{2,0}-3_{1,3}$) (top),
CH$_3$OH$(7_{5,3}-6_{4,2})$ (middle), and
CH$_3$OH$(14_{1,13}-14_{0,14})$ (bottom) lines, respectively. The
grey-scale shows the 860\,$\mu$m continuum emission in all panels. The
CH$_3$OH($2_{2,0}-3_{1,3}$) map is contoured positive from 0.46 to
3.68\,Jy (step 0.46\,Jy), and negative from -0.46 to -1.84\,Jy (step
-0.46\,Jy). The CH$_3$OH$(7_{5,3}-6_{4,2})$ date are contoured
positive from 0.6 to 4.8\,Jy in 0.6\,Jy steps, and negative at $-0.6$
\& $-1.2$\,Jy. The CH$_3$OH$(14_{1,13}-14_{0,14})$ are contoured
positive from 0.76 to 5.32\,Jy in 0.76\,Jy steps and negative at
$-0.76$ and $-1.52$\,Jy. The synthesized beams are shown at the
bottom-left of each panel (full ellipse for continuum and lined
ellipse for CH$_3$OH).}
\label{ch3oh_channel}
\end{figure}

The previously observed CH$_3$OH$(5_{1,4}-4_{2,2})$ line at
216.946\,GHz peaked toward the mm continuum and exhibited a velocity
gradient in the east-west direction (Fig.~3 in
\citealt{beuther2004a}). Figure \ref{ch3oh_channel} presents channel
maps for the CH$_3$OH($2_{2,0}-3_{1,3}$), CH$_3$OH$(7_{5,3}-6_{4,2})$,
and CH$_3$OH$(14_{1,13}-14_{0,14})$ lines at 340.141, 338.722, and
349.107\,GHz, respectively. The spatial distribution of the three
lines is not exactly the same: the CH$_3$OH($2_{2,0}-3_{1,3}$) and
CH$_3$OH$(14_{1,13}-14_{0,14})$ both show red emission to the west of
the continuum peak but only weak or barely detectable blue emission to
the east (Fig.~\ref{ch3oh_channel}\,top and bottom). Contrary to this
we cannot discern such a signature in the CH$_3$OH$(7_{5,3}-6_{4,2})$
line, the emission peaks north and south of the submm continuum peak
and seems to be affected by the outflow in that direction
(Fig.~\ref{ch3oh_channel}\,middle). While none of the lines shows a
clear east-west velocity gradient indicative of the proposed disk as
suggested by the previous CH$_3$OH$(5_{1,4}-4_{2,2})$ observations,
the red emission of the CH$_3$OH($2_{2,0}-3_{1,3}$) and
CH$_3$OH$(14_{1,13}-14_{0,14})$ lines are at least consistent with it.

Since CH$_3$OH is a slightly asymmetric rotor with a very rich
spectrum in the mm and submm bands (see Table~\ref{lines} or
\citealt{leurini2004} for a more detailed CH$_3$OH discussion), it is
interesting to investigate in what kind of environment one finds the
molecule. As shown in Table~\ref{lines}, within the given bandpass we
observe CH$_3$OH lines spanning upper level energy ranges between 44
and 695\,K. Since the lower energy CH$_3$OH($2_{2,0}-3_{1,3}$)
($E_{\rm{upper}}=44$\,K) and the higher energy
CH$_3$OH$(14_{1,13}-14_{0,14})$ ($E_{\rm{upper}}=259$\,K) lines are
more confined to the inner center than the intermediate energy
CH$_3$OH$(7_{5,3}-6_{4,2})$ line ($E_{\rm{upper}}=87$\,K), one could
speculate that the temperatures within the outflow might be in the
latter regime and thus influence this line most strongly. However, not
all lines of the CH$_3$OH$(7-6)$ series resemble the
outflow-influenced morphology, but others also trace the inner core
again (e.g., CH$_3$OH$(7_{4,3}-6_{4,3})$). In the literature one finds
cases where CH$_3$OH traces the protostellar cores (e.g.,
\citealt{wyrowski1999,vandertak2000}) as well as molecular outflows
(e.g., \citealt{bachiller1998,beuther2002d,joergensen2004}), and
currently it is not well understood which CH$_3$OH line is suitable to
trace one or the other effect. Nevertheless, because CH$_3$OH is a
molecule with often more than one line in a given spectral setup, it
is a potentially useful tool to investigate the different
processes. Since the CH$_3$OH collisional rates were calculated
recently \citep{pottage2004}, the potential of CH$_3$OH as an
interstellar tracer is likely going to be exploited more
extensively than before (e.g., \citealt{leurini2004}).

\section{Discussion}

\subsection{Continuum emission}
\label{discussion_cont}

It is interesting to see that even at this high spatial resolution
IRAS\,18089-1732 still does not split up in any sub-fragments. The
3$\sigma$ continuum sensitivity of 105\,mJy corresponds to a mass
sensitivity of 0.3, 0.7, or 3.0\,M$_{\odot}$, depending on the assumed
temperature which we set to 350\,K (the derived CH$_3$CN hot core
temperature), 100\,K (a more typical hot core temperature), and 30\,K
(which should be the temperature of the surrounding envelope),
respectively. Since potential lower-mass companions of an evolving
massive cluster should be distributed a little bit further out from
the central core (e.g., \citealt{lada2003}), lower temperature values
of 100 and/or 30\,K appear to be a more realistic for potential
companions. Nevertheless, we find no companion down to a $3\sigma$
mass limit of 0.7 and 3\,M$_{\odot}$ at 100 and 30\,K,
respectively. The non-detection of any companion source supports the
previous reasoning that the SiO outflow is indeed emanating from the
main massive submm continuum source and not caused by a lower-mass
cluster member \citep{beuther2004b}. The small offset of the SiO
outflow axis from the submm peak position ($\sim 1''$,
\citealt{beuther2004b}) is still a bit puzzling, but it can 
likely be attributed to the underlying flow geometry and the resulting
shock-induced local SiO enhancements (e.g.,
\citealt{schilke1997a}).

The single submm continuum peak is also distinctively different to the
recent study by \citet{beuther2004c} who found the HMPO
IRAS\,19410+2336 to split up into a cluster of sources with a
protocluster mass function resembling the stellar Initial Mass
Function. Contrary to this, there exist other prominent examples which
do not split up into clusters, e.g., IRAS\,20126+4104, but remain one
mm continuum source down to the highest spatial resolution
\citep{cesaroni1999}\footnote{It is noteworthy that the mm continuum
peak in IRAS\,20126+4104 contains a ionized double jet indicating a
binary system \citep{hofner1999}. This may be responsible for the
large-scale precession of the molecular outflow
\citep{shepherd2000,cesaroni2005}.}, similar to the case of
IRAS\,18089-1732.

What is causing these observational differences? On the one hand, it
is possible that the study by \citet{beuther2004c} had very favorable
observational circumstances, because they observed IRAS\,19410+2336 in
three configurations with the Plateau de Bure Interferometer and thus
had an exceptionally good uv-coverage sampling various spatial
scales. Our SMA data have been observed with baselines not shorter
than 20\,m and thus does not trace the corresponding larger spatial
scales, especially at this high frequencies. Excluding the compact
Plateau de Bure configuration, \citet{beuther2004c} would not have
detected that many sub-sources in IRAS\,19410+2336 as well. Therefore,
observational differences might account to some degree for the
discussed differences. On the other hand, could it be possible that
some massive star-forming regions evolve in a fragmenting cluster-like
way like IRAS\,19410+2336, whereas other star-forming regions exhibit
the core-halo structure like IRAS\,18089-1732 at the earliest
evolutionary stages and develop the cluster-like structure at later
evolutionary stages? This scenario would imply that at least for such
kind of sources the massive cluster members would form first and the
low- and intermediate-mass members later. One way to solve this
question is to choose a number of similar sources (similar in the
sense of evolutionary stage, luminosity and distance; IRAS\,20126+4104
and IRAS\,19410+2336 for example fulfill these criteria) and observe
all these sources with exactly the same interferometric configurations
to trace the same spatial scales. If this experiment would still
result in the same observational differences as encountered here, this
would be a hint to the existence of a variety of evolutionary paths
for massive star-forming regions.

Another result from the continuum study is that it is difficult to
derive a consistent intensity distribution and thus density
distribution on this small spatial scales because apparently
IRAS\,18089-1732 is not symmetric at all. It is likely that the
central core still harbors multiple unresolved sources which could
cause these asymmetries. In any regard, the usual approach of circular
averaging and then deriving density distributions which are
interpreted within one or the other theoretical model has to be
approached cautiously and needs to be re-checked for each studied
source separately.

\subsection{Disks in massive star formation}

We are able to confirm the previously suggested disk/outflow scenario
for the HMPO IRAS\,18089-1732 at higher spatial resolution. The
HCOOCH$_3$ velocity shift is approximately perpendicular to the SiO
outflow axis. However, even at this spatial resolution it is hard to
judge whether we really trace a Keplerian accretion disk (Keplerian in
the sense of a Keplerian velocity profile dominated by the central
protostar) or rather a rotating larger-scale envelope which rotates
perpendicular to the outflow but which might not be Keplerian at
all. The fact that a large fraction of the central gas core appears to
be associated with the rotation indicates that it might not be in
Keplerian motion. This implies a follow-up question whether massive
accretion disks need to be Keplerian like their low-mass counterparts,
or whether massive disks may be non-Keplerian, potentially
self-gravitating structures (self-gravitating in the sense that the
gravitational potential of the disk-star system is dominated by the
disk). While the observations toward prototypical high-mass disk
source IRAS\,20126+4104 favor a Keplerian disk, although with a
central mass of $\sim 7$\,M$_{\odot}$ it is not a very massive source
(e.g.,
\citealt{cesaroni2005}), other sources indicate disk masses well in
excess of the central source (e.g.,
\citealt{sandell2003,chini2004}). For IRAS\,18089-1732, we find a rotating
structure which is consistent with an accretion disk/rotating
envelope, but unfortunately the data still lack the spatial resolution
to study the physical structure of the disk. Follow-up observations at
higher angular resolution are one tool to tackle these questions.

In addition, the question posed in \S\ref{intro} to find the most
suitable disk-tracing molecules has not been sufficiently
answered. The previously successful molecule CH$_3$CN has proved in
this source as not suitable.  It is not clear whether simply the
chosen K-ladder line series is a bad one because the opacity $\tau$
scales with the frequency $\nu$ like:

\begin{eqnarray}
\tau & = & \frac{c^2}{8\pi^2}\left(\frac{A}{\nu^3}\right)\left(\frac{\nu}{\Delta \nu}\right)\left(e^{\frac{h\nu}{kT}}-1\right)N_u\\
\Leftrightarrow \tau & = & constant \times \nu\left(e^{\frac{h\nu}{kT}}-1\right)N_u
\end{eqnarray}

with A the Einstein coefficient, $\Delta \nu$ the line width, and
$N_u$ the column density of the upper state.  Using the temperature
estimate of 350\,K and assuming the same line width for all CH$_3$CN
line series, we can approximate the change in optical depth for the
CH$_3$CN line series in the 349\,GHz and 220\,GHz bands by:

\begin{eqnarray}
\frac{\tau_{CH_3CN(19-18)}}{\tau_{CH_3CN(12-11)}}\sim 1.6\times \frac{349}{220}\times \frac{N_{u,CH_3CN(19-18)}}{N_{u,CH_3CN(12-11)}}
\end{eqnarray}

In case the CH$_3$CN column densities in both transitions are similar,
which is given to first order in thermal equilibrium at the estimated
temperature of 350\,K, the line opacity in the 860\,$\mu$m band would
be approximately 2.5 times larger than in the 1\,mm band. Therefore,
higher optical depth at 860\,$\mu$m could contribute to the
observation that we do not find a disk signature in the
CH$_3$CN$(19_k-18_k)$ transitions. However, it is possible that already the
1\,mm lines are optically thick which has to be checked via future
observations of the latter line series.

Contrary to this, the HCOOH$_3$ molecule traces the disk-like
structure in the 1\,mm and the 860\,$\mu$m band successfully and
apparently does not suffer from too high opacity. However, in spite of
its ubiquity in many spectral bands
\citep{sutton1985,schilke1997b,schilke2001}, spectral HCOOCH$_3$ lines
are often rather weak and and blended with various components, and
therefore observationally more difficult to tackle. To start
statistically significant studies of massive accretion disks, we still
have to find a suitable molecular transition which unambiguously can
trace disks and does not require a too large amount of observing time.

In spite of these problems, it is promising that more and more massive
disk candidates are being studied. With concerted efforts from various
groups employing all available interferometers, it is likely that we
will get a much better understanding of massive disks over the next
decade.

\section{Conclusions and Summary}

Fifty spectral lines from eighteen different species spanning an
energy range from 17 to 747\,K were detected within the 4\,GHz
bandpass at 860\,$\mu$m. While the previously assumed best disk tracer
CH$_3$CN in this band is optically thick and thus not suitable for
follow-up disk studies in this source, HCOOCH$_3$ appears optically
thin and shows a velocity-disk signature similar to well-known
low-mass disk sources. The estimated mass of this rotating structure,
assuming equilibrium between the gravitational and centrifugal forces,
is 16/(sin$^2(i)$)\,M$_{\odot}$, of the same order as the core mass
derived from the submm dust continuum emission. This implies that a
significant fraction of the central core mass is associated with the
rotation, and the assumed disk-like structure may not be
Keplerian. Within this mini-line-survey, HCOOCH$_3$ appears as one of
the best disk-tracing molecules.

Although CH$_3$CN$(19_k-18_k)$ is optically thick, we can use the
K-ladder to estimate a temperature in the region. We find a
temperature of approximately 350\,K, consistent with previous
HCOOCH$_3$ estimates \citep{beuther2004a}, and a hot molecular core
has already formed.

Even at $1''$ spatial resolution, we do not detect any companion
source in the field of view down to mass limits between 0.2 and
3\,M$_{\odot}$ (depending on the assumed temperature). Since this
finding is distinctively different to the recent study toward the HMPO
IRAS\,19410+2336 it might indicate different evolutionary paths for
the formation of massive clusters. However, observational systematics
can also account for some of the differences, and more detailed
investigations of various sources with the same observational settings
are necessary to draw firm conclusions. Furthermore, the central dust
core is highly non-circular, indicating potential difficulties for
circular averaged studies of intensity profiles at this high spatial
resolution.

\acknowledgments{We like to thank Todd Hunter for the fruitful
discussions about this source. Thanks a lot also to Peter Schilke for
providing the XCLASS software. Furthermore, we appreciate the
constructive and quick report from an anonymous
referee. H.B. acknowledges financial support by the
Emmy-Noether-Program of the Deutsche Forschungsgemeinschaft (DFG,
grant BE2578/1).}

%\bibliography{/home/hbeuther/bibliography}   
%\bibliography{/Users/henrikbeuther/paper/bibliography}

\begin{thebibliography}{41}
\expandafter\ifx\csname natexlab\endcsname\relax\def\natexlab#1{#1}\fi

\bibitem[{{Bachiller} {et~al.}(1998){Bachiller}, {Codella}, {Colomer},
  {Liechti}, \& {Walmsley}}]{bachiller1998}
{Bachiller}, R., {Codella}, C., {Colomer}, F., {Liechti}, S., \& {Walmsley},
  C.~M. 1998, \aap, 335, 266

\bibitem[{{Beltr{\' a}n} {et~al.}(2004){Beltr{\' a}n}, {Cesaroni}, {Neri},
  {Codella}, {Furuya}, {Testi}, \& {Olmi}}]{beltran2004}
{Beltr{\' a}n}, M.~T., {Cesaroni}, R., {Neri}, R., {et~al.} 2004, \apjl, 601,
  L187

\bibitem[{{Beuther} {et~al.}(2004{\natexlab{a}}){Beuther}, {Hunter}, {Zhang},
  {Sridharan}, {Zhao}, {Sollins}, {Ho}, {Ohashi}, {Su}, {Lim}, \&
  {Liu}}]{beuther2004b}
{Beuther}, H., {Hunter}, T.~R., {Zhang}, Q., {et~al.} 2004{\natexlab{a}},
  \apjl, 616, L23

\bibitem[{{Beuther} \& {Schilke}(2004)}]{beuther2004c}
{Beuther}, H. \& {Schilke}, P. 2004, Science, 303, 1167

\bibitem[{{Beuther} {et~al.}(2002{\natexlab{a}}){Beuther}, {Schilke}, {Gueth},
  {McCaughrean}, {Andersen}, {Sridharan}, \& {Menten}}]{beuther2002d}
{Beuther}, H., {Schilke}, P., {Gueth}, F., {et~al.} 2002{\natexlab{a}}, \aap,
  387, 931

\bibitem[{{Beuther} {et~al.}(2002{\natexlab{b}}){Beuther}, {Schilke}, {Menten},
  {Motte}, {Sridharan}, \& {Wyrowski}}]{beuther2002a}
{Beuther}, H., {Schilke}, P., {Menten}, K.~M., {et~al.} 2002{\natexlab{b}},
  \apj, 566, 945

\bibitem[{{Beuther} {et~al.}(2002{\natexlab{c}}){Beuther}, {Schilke},
  {Sridharan}, {Menten}, {Walmsley}, \& {Wyrowski}}]{beuther2002b}
{Beuther}, H., {Schilke}, P., {Sridharan}, T.~K., {et~al.} 2002{\natexlab{c}},
  \aap, 383, 892

\bibitem[{{Beuther} {et~al.}(2002{\natexlab{d}}){Beuther}, {Walsh}, {Schilke},
  {Sridharan}, {Menten}, \& {Wyrowski}}]{beuther2002c}
{Beuther}, H., {Walsh}, A., {Schilke}, P., {et~al.} 2002{\natexlab{d}}, \aap,
  390, 289

\bibitem[{{Beuther} {et~al.}(2004{\natexlab{b}}){Beuther}, {Zhang}, {Hunter},
  {Sridharan}, {Zhao}, {Sollins}, {Ho}, {Liu}, {Ohashi}, {Su}, \&
  {Lim}}]{beuther2004a}
{Beuther}, H., {Zhang}, Q., {Hunter}, T.~R., {et~al.} 2004{\natexlab{b}},
  \apjl, 616, L19

\bibitem[{{Cesaroni} {et~al.}(1999){Cesaroni}, {Felli}, {Jenness}, {Neri},
  {Olmi}, {Robberto}, {Testi}, \& {Walmsley}}]{cesaroni1999}
{Cesaroni}, R., {Felli}, M., {Jenness}, T., {et~al.} 1999, \aap, 345, 949

\bibitem[{{Cesaroni} {et~al.}(1997){Cesaroni}, {Felli}, {Testi}, {Walmsley}, \&
  {Olmi}}]{cesaroni1997}
{Cesaroni}, R., {Felli}, M., {Testi}, L., {Walmsley}, C.~M., \& {Olmi}, L.
  1997, \aap, 325, 725

\bibitem[{{Cesaroni} {et~al.}(2005){Cesaroni}, {Neri}, {Olmi}, {Walmsley}, \&
  {Hofner}}]{cesaroni2005}
{Cesaroni}, R., {Neri}, R., {Olmi}, L., {Walmsley}, C., \& {Hofner}, P. 2005,
  \aap~accepted

\bibitem[{{Chini} {et~al.}(2004){Chini}, {Hoffmeister}, {Kimeswenger},
  {Nielbock}, {N{\" u}rnberger}, {Schmidtobreick}, \& {Sterzik}}]{chini2004}
{Chini}, R., {Hoffmeister}, V., {Kimeswenger}, S., {et~al.} 2004, \nat, 429,
  155

\bibitem[{{Davis} {et~al.}(2004){Davis}, {Varricatt}, {Todd}, \& {Ramsay
  Howat}}]{davis2004}
{Davis}, C.~J., {Varricatt}, W.~P., {Todd}, S.~P., \& {Ramsay Howat}, S.~K.
  2004, \aap, 425, 981

\bibitem[{{Garay} {et~al.}(2004){Garay}, {Fa{\' u}ndez}, {Mardones},
  {Bronfman}, {Chini}, \& {Nyman}}]{garay2004}
{Garay}, G., {Fa{\' u}ndez}, S., {Mardones}, D., {et~al.} 2004, \apj, 610, 313

\bibitem[{{Gibb} {et~al.}(2003){Gibb}, {Hoare}, {Little}, \&
  {Wright}}]{gibb2003}
{Gibb}, A.~G., {Hoare}, M.~G., {Little}, L.~T., \& {Wright}, M.~C.~H. 2003,
  \mnras, 339, 1011

\bibitem[{{Guilloteau} \& {Dutrey}(1998)}]{guilloteau1998}
{Guilloteau}, S. \& {Dutrey}, A. 1998, \aap, 339, 467

\bibitem[{{Hildebrand}(1983)}]{hildebrand1983}
{Hildebrand}, R.~H. 1983, \qjras, 24, 267

\bibitem[{{Hofner} {et~al.}(1999){Hofner}, {Cesaroni}, {Rodr{\'{\i}}guez}, \&
  {Mart{\'{\i}}}}]{hofner1999}
{Hofner}, P., {Cesaroni}, R., {Rodr{\'{\i}}guez}, L.~F., \& {Mart{\'{\i}}}, J.
  1999, \aap, 345, L43

\bibitem[{{J{\o}rgensen} {et~al.}(2004){J{\o}rgensen}, {Hogerheijde}, {Blake},
  {van Dishoeck}, {Mundy}, \& {Sch{\" o}ier}}]{joergensen2004}
{J{\o}rgensen}, J.~K., {Hogerheijde}, M.~R., {Blake}, G.~A., {et~al.} 2004,
  \aap, 415, 1021

\bibitem[{{Lada} \& {Lada}(2003)}]{lada2003}
{Lada}, C.~J. \& {Lada}, E.~A. 2003, \araa, 41, 57

\bibitem[{{Leurini} {et~al.}(2004){Leurini}, {Schilke}, {Menten}, {Flower},
  {Pottage}, \& {Xu}}]{leurini2004}
{Leurini}, S., {Schilke}, P., {Menten}, K.~M., {et~al.} 2004, \aap, 422, 573

\bibitem[{{M{\" u}ller} {et~al.}(2001){M{\" u}ller}, {Thorwirth}, {Roth}, \&
  {Winnewisser}}]{mueller2001}
{M{\" u}ller}, H.~S.~P., {Thorwirth}, S., {Roth}, D.~A., \& {Winnewisser}, G.
  2001, \aap, 370, L49

\bibitem[{{Pestalozzi} {et~al.}(2004){Pestalozzi}, {Elitzur}, {Conway}, \&
  {Booth}}]{pestalozzi2004}
{Pestalozzi}, M.~R., {Elitzur}, M., {Conway}, J.~E., \& {Booth}, R.~S. 2004,
  \apjl, 603, L113

\bibitem[{{Pottage} {et~al.}(2004){Pottage}, {Flower}, \&
  {Davis}}]{pottage2004}
{Pottage}, J.~T., {Flower}, D.~R., \& {Davis}, S.~L. 2004, \mnras, 352, 39

\bibitem[{{Poynter} \& {Pickett}(1985)}]{poynter1985}
{Poynter}, R.~L. \& {Pickett}, H.~M. 1985, \ao, 24, 2235

\bibitem[{{Sandell} {et~al.}(2003){Sandell}, {Wright}, \&
  {Forster}}]{sandell2003}
{Sandell}, G., {Wright}, M., \& {Forster}, J.~R. 2003, \apjl, 590, L45

\bibitem[{{Schilke} {et~al.}(2001){Schilke}, {Benford}, {Hunter}, {Lis}, \&
  {Phillips}}]{schilke2001}
{Schilke}, P., {Benford}, D.~J., {Hunter}, T.~R., {Lis}, D.~C., \& {Phillips},
  T.~G. 2001, \apjs, 132, 281

\bibitem[{{Schilke} {et~al.}(1997{\natexlab{a}}){Schilke}, {Groesbeck},
  {Blake}, \& {Phillips}}]{schilke1997b}
{Schilke}, P., {Groesbeck}, T.~D., {Blake}, G.~A., \& {Phillips}, T.~G.
  1997{\natexlab{a}}, \apjs, 108, 301

\bibitem[{{Schilke} {et~al.}(1997{\natexlab{b}}){Schilke}, {Walmsley}, {Pineau
  des Forets}, \& {Flower}}]{schilke1997a}
{Schilke}, P., {Walmsley}, C.~M., {Pineau des Forets}, G., \& {Flower}, D.~R.
  1997{\natexlab{b}}, \aap, 321, 293

\bibitem[{{Scoville} {et~al.}(1993){Scoville}, {Carlstrom}, {Chandler},
  {Phillips}, {Scott}, {Tilanus}, \& {Wang}}]{scoville1993}
{Scoville}, N.~Z., {Carlstrom}, J.~E., {Chandler}, C.~J., {et~al.} 1993, \pasp,
  105, 1482

\bibitem[{{Shepherd} {et~al.}(2000){Shepherd}, {Yu}, {Bally}, \&
  {Testi}}]{shepherd2000}
{Shepherd}, D.~S., {Yu}, K.~C., {Bally}, J., \& {Testi}, L. 2000, \apj, 535,
  833

\bibitem[{{Sridharan} {et~al.}(2002){Sridharan}, {Beuther}, {Schilke},
  {Menten}, \& {Wyrowski}}]{sridha}
{Sridharan}, T.~K., {Beuther}, H., {Schilke}, P., {Menten}, K.~M., \&
  {Wyrowski}, F. 2002, \apj, 566, 931

\bibitem[{{Su} {et~al.}(2004){Su}, {Liu}, {Lim}, {Ohashi}, {Beuther}, {Zhang},
  {Sollins}, {Hunter}, T.K., {Zhao}, \& {Ho}}]{su2004b}
{Su}, Y.-N., {Liu}, S.-Y., {Lim}, J., {et~al.} 2004, \apjl, 616, L39

\bibitem[{{Sutton} {et~al.}(1985){Sutton}, {Blake}, {Masson}, \&
  {Phillips}}]{sutton1985}
{Sutton}, E.~C., {Blake}, G.~A., {Masson}, C.~R., \& {Phillips}, T.~G. 1985,
  \apjs, 58, 341

\bibitem[{{Torrelles} {et~al.}(1996){Torrelles}, {Gomez}, {Rodriguez},
  {Curiel}, {Ho}, \& {Garay}}]{torrelles1996}
{Torrelles}, J.~M., {Gomez}, J.~F., {Rodriguez}, L.~F., {et~al.} 1996, \apjl,
  457, L107+

\bibitem[{{van der Tak} {et~al.}(2000){van der Tak}, {van Dishoeck}, \&
  {Caselli}}]{vandertak2000}
{van der Tak}, F.~F.~S., {van Dishoeck}, E.~F., \& {Caselli}, P. 2000, \aap,
  361, 327

\bibitem[{{Williams} {et~al.}(2004){Williams}, {Fuller}, \&
  {Sridharan}}]{williams2004}
{Williams}, S.~J., {Fuller}, G.~A., \& {Sridharan}, T.~K. 2004, \aap, 417, 115

\bibitem[{{Wyrowski} {et~al.}(1999){Wyrowski}, {Schilke}, {Walmsley}, \&
  {Menten}}]{wyrowski1999}
{Wyrowski}, F., {Schilke}, P., {Walmsley}, C.~M., \& {Menten}, K.~M. 1999,
  \apjl, 514, L43

\bibitem[{{Zhang} {et~al.}(1998){Zhang}, {Hunter}, \& {Sridharan}}]{zhang1998a}
{Zhang}, Q., {Hunter}, T.~R., \& {Sridharan}, T.~K. 1998, \apjl, 505, L151

\bibitem[{{Zhang} {et~al.}(2002){Zhang}, {Hunter}, {Sridharan}, \&
  {Ho}}]{zhang2002}
{Zhang}, Q., {Hunter}, T.~R., {Sridharan}, T.~K., \& {Ho}, P.~T.~P. 2002, \apj,
  566, 982

\end{thebibliography}
%\bibliographystyle{aa}    % this does the style, aa.bst necessary

\begin{table}
\caption{Submillimeter Continuum data \label{submm}}
\begin{center}
\begin{tabular}{lr}
\hline
$S_{\rm{peak}}$ & 1.400\,mJy \\
$S_{\rm{int}}$  & 6440\,mJy \\
$M$ at 100\,K  & 45\,M$_{\odot}$\\
$N$ at 100\,K  & $3.5\times 10^{23}$\,cm$^{-2}$ \\
$M$ at 350\,K  & 12\,M$_{\odot}$\\
$N$ at 350\,K  & $9.5\times 10^{23}$\,cm$^{-2}$ \\
\hline
\end{tabular}
\end{center}
\end{table}

\begin{deluxetable}{lrrr}
\tablecaption{Observed lines \label{lines}}
\tablewidth{0pt}
\tablehead{
\colhead{$\nu$} & \colhead{line} & \colhead{$S_{\rm{peak}}^b$} & \colhead{$E_{\rm{upper}}$} \\
\colhead{GHz} & \colhead{} & \colhead{[K]} & \colhead{[K]}
}
\startdata
LSB\\
\hline
338.5041 & CH$_3$OH($7_{2,6}-6_{1,5}$)E & 5.0 & 153 \\
338.5126 & CH$_3$OH($7_{6,2}-6_{5,1}$)A & 4.1 & 145 \\
338.5303 & CH$_3$OH($7_{6,2}-6_{5,1}$)E & 3.7 & 160 \\
338.5408 & CH$_3$OH($7_{5,2}-6_{5,2}$)A$^+$$^a$ & 4.9 & 115 \\
338.5432 & CH$_3$OH($7_{5,2}-6_{5,2}$)A$^-$$^a$ &  & 115 \\
338.5599 & CH$_3$OH($7_{2,5}-6_{2,5}$)E & 5.8 & 128 \\
338.5832 & CH$_3$OH($7_{5,2}-6_{5,2}$)E & 5.3 & 113 \\
338.6150 & CH$_3$OH($7_{4,3}-6_{4,3}$)E & 5.0 & 86 \\
338.6399 & CH$_3$OH($7_{5,3}-6_{4,2}$)A$^+$ & 4.9 & 103 \\
338.7216 & CH$_3$OH($7_{5,3}-6_{4,2}$)E$^a$ & 4.8 & 87 \\
338.7229 & CH$_3$OH($7_{3,5}-6_{2,4}$)E$^a$ & & 91 \\
338.7604 & $^{13}$CH$_3$OH($13_{7,7}-12_{7,6}$)A$^+$ & 2.2 & 206 \\
338.7858 & $^{34}$SO$_2$($14_{4,10}-12_{3,1}$) & -- & 134 \\             
338.8862 & C$_2$H$_5$OH($15_{7,8}-15_{6,9}$)$^a$ & -- & 162 \\
338.8873 & C$_2$H$_5$OH($15_{7,9}-15_{6,10}$)$^a$ & -- & 162 \\
339.3126 & C$_2$H$_5$OH($12_{7,5}-12_{6,6}$) & -- & 126 \\
339.3415 & SO($3_3-2_3$) & 1.6 & 25 \\
339.4917 & CH$_3$OCH$_3$($19_{1,18}-18_{2,17}$)AA & 3.4 & 176 \\
339.8576 & $^{34}$SO($8_9-7_8$) & 2.6 & 77 \\
339.9041 & NH$_2$CHO($16_{6,11}-15_{6,10}$) & -- & 246 \\
340.0527 & C$^{33}$S($7-6$) & 2.6 & 65 \\
340.1412 & CH$_3$OH($2_{2,0}-3_{1,3}$)A$^+$  & 4.4 & 44 \\
340.1893 & C$_2$H$_5$OH($6_{5,2}-5_{4,1}$) & 1.6 & 49 \\
340.2486 & CN($3,\frac{7}{2},\frac{5}{2}-2,\frac{5}{2},\frac{3}{2}$)&--& 32\\
340.3164 & SO$_2$($28_{2,26}-28_{1,27}$) & 1.6 & 392 \\
340.3937 & CH$_3$OH($16_{6,10}-17_{5,13}$)A$^+$ & 3.4 & 508 \\
340.4492 & OCS($28-27$) 2.3 & 237  \\
\hline
USB \\
\hline
348.5319 &  H$_2$CS($10_{1,9}-9_{1,8}$) & 4.3 & 105 \\
348.9095 &  HCOOCH$_3$($28_{9,20}-27_{9,19}$)E$^{a}$ & 1.7 & 294 \\
348.9114 & CH$_3$CN($19_9-18_9$)$^{a}$ & & 745 \\
348.9150 & HCOOCH$_3$($28_{9,20}-27_{9,19}$)A$^{a}$ & & 294 \\
349.0485 & HCOOCH$_3$($28_{9,19}-27_{9,18}$)E & -- & 294 \\
349.1070 & CH$_3$OH($14_{1,13}-14_{0,14}$)A & 6.2 & 259 \\
349.1253 & CH$_3$CN($19_7-18_7$) & 2.1 & 517 \\
349.1732 & CH$_3$$^{13}$CN($19_4-18_4$) & -- & 281 \\
349.2121 & CH$_3$CN($19_6-18_6$) & 3.8 & 424 \\
349.2858 & CH$_3$CN($19_5-18_5$) & 3.2 & 346 \\
349.3379 & CCH($4,\frac{9}{2},\frac{9}{2}-3,\frac{7}{2},\frac{7}{2}$)&--&41\\
349.3461 & CH$_3$CN($19_4-18_4$) & 3.4 & 281 \\
349.3930 & CH$_3$CN($19_3-18_3$) & 3.4 & 231 \\
349.4266 & CH$_3$CN($19_2-18_2$) & 4.8 & 196 \\
349.4467 & CH$_3$CN($19_1-18_1$)$^{a}$ &     & 175 \\
349.4534 & CH$_3$CN($19_0-18_0$)$^{a}$ & 6.1 & 167 \\
350.1031 & $^{13}$CH$_3$OH($1_{1,0}-0_{0,0}$)A$^+$ & -- & 17 \\
350.2877 & CH$_3$OH($15_{3,13}-16_{4,13}$) & 2.0 & 695 \\
350.3333 & HNCO($16_{1,16}-15_{1,15}$) & 3.0 & 185 \\
350.4235 & CH$_3$CN($v_8=1$)($19_{-2}-18_{-2}$) & 2.1 & 747 \\
350.4423 & HCOOCH$_3$($28_{8,20}-27_{8,20}$)E$^{a}$ & 2.1 & 283 \\
350.4449 & CH$_3$CN($v_8=1$)($19_0-18_0$)$^{a}$ & & 693
\enddata
%% Text for table notes should follow after the \enddata but before
%% the \end{deluxetable}. Make sure there is at least one \tablenotemark
%% in the table for each \tablenotetext.
\tablenotetext{a}{\footnotesize line-blend}
\tablenotetext{b}{\footnotesize The peak intensities $S_{\rm{peak}}$
were measured toward the submm continuum peak position after imaging
the whole data-cube with 2\,km/s spectral resolution. Lines which were
too weak to image are marked with a ``--'', and for line-blends we
just list the intensity for one component.}
\end{deluxetable}

\begin{deluxetable}{lr}
\tablecaption{Approximate area within 50\% contours$^a$ \label{area}}
\tablewidth{0pt}
\tablehead{
\colhead{} & \colhead{Area} \\
\colhead{} & \colhead{[arcsec$^2$]} 
}
\startdata
860\,$\mu$m continuum & 1.0 \\
HCOOCH$_3$($28_{9,20}-27_{9,19}$)E & 1.2 \\
CH$_3$CN($19_2-18_2$) & 1.8\\
CH$_3$CN($v_8=1$)($19_{-2}-18_{-2}$) & 1.6\\
CH$_3$OH($14_{1,13}-14_{0,14}$)A & 1.7\\
$^{13}$CH$_3$OH($13_{7,7}-12_{7,6}$)A$^+$ & 1.3\\
CH$_3$OCH$_3$($19_{1,18}-18_{2,17}$)AA & 1.9\\
H$_2$CS($10_{1,9}-9_{1,8}$) & 2.7\\
HNCO($16_{1,16}-15_{1,15}$) & 1.6\\
OCS($28-27$) & 3.8 \\
C$^{33}$S(7--6) & 2.3
\enddata
\tablenotetext{a}{\footnotesize The area is calculated assuming an 
elliptical source geometry and measuring the major and minor axes of 
the 50\% contour levels.}
\end{deluxetable}

\end{document}